\documentclass[12pt]{article}
\usepackage{epsf}
\usepackage{amsmath}

\usepackage{graphics}
\usepackage{cite}

\begin{titlepage}

\begin{document}

\hfill January 2023

\begin{center}

{\bf \LARGE An $SU(15)$ Approach to Bifermion Classification}\\
\vspace{0.5cm}
{\bf Claudio Corian\`o}\footnote{claudio.coriano@le.infn.it}\\
{\bf Paul H. Frampton}\footnote{paul.h.frampton@gmail.com}\\
{\bf Dario Melle}\footnote{dario.melle@studenti.unisalento.it}\\
\vspace{0.5cm}
{\it Dipartimento di Matematica e Fisica ``Ennio De Giorgi",\\ 
Universit\`{a} del Salento and INFN-Lecce,\\ Via Arnesano, 73100 Lecce, Italy\\
National Center for HPC, Big Data and Quantum Computing}

\vspace{0.2in}
{\bf Thomas W. Kephart}\footnote{tom.kephart@gmail.com}\\
\vspace{0.5cm}
{\it Department of Physics and Astronomy, Vanderbilt University,\\
Nashville, TN 37235, USA.}

\vspace{0.2in}
{\bf Tzu-Chiang Yuan}\footnote{tcyuan@phys.sinica.edu.tw}\\
\vspace{0.5cm}
{\it Institute of Physics, Academia Sinica, Nangang, Taipei 11529, Taiwan.}

\end{center}

\begin{abstract}
\noindent
One interesting way to extend the standard model is the hypothesis
of bifermions which are bosons which couple to pairs of
quarks and leptons.
We point out that $SU(15)$ grand unification gives a natural
way to classify bifermions and discuss
leptoquarks, biquarks and bileptons.
In fact,  $SU(15)$ provides an ideal covering group as it contains all possible
bifermions within a single model.
\end{abstract}

\end{titlepage}

\noindent
The standard model (SM) of particle theory has remained robust
and only occasionally tantalising hints have appeared from experiment 
about how
to extend it. If and when these hints have become more 
definite they
are likely to influence all of theoretical physics by clarifying
the choices which Nature has made. A recent disappointment
was that the anomalies in B decays which had stubbornly remained
for the eight years
2014-2022 at the $3\sigma$ level have now been withdrawn\cite{LHCb}.
The present article is intended to be useful for the time when further
discrepancies from the standard model appear.
\noindent
One attempt at grand unification\cite{FL} involves the gauge group $SU(15)$
where all 15 states of a quark-lepton family are in the defining representation
and every possible leptoquark is present in the adjoint representation 
which provides a useful classification.
\noindent
The adjoint appears in $15 \times 15^* = 1 + 224$ and contains 72 leptoquarks which transform
in irreducible representations of the standard model gauge group\\
$(SU(3)_C, SU(2)_L)_Y$  with $Q=T_3 + Y/2$ in four sets of 18 as follows

\begin{equation}
B = +1/3, L=+1, ~~~~~2(3, 2)_{-5/3}~~~	Q=(-1/3,-4/3)	~~~ue^-,de^-     \nonumber \\
\end{equation}
\begin{equation}
\hspace{1.3in}~~~(3, 2)_{1/3} ~~~~~~	Q=(2/3. -1/3)	~~~ u\nu,d\nu     \nonumber \\
\end{equation}

\bigskip

\begin{equation}
B = -1/3, L=+1, ~~~~~~~~~~~~~~~~~~~2(3^*,1)_{-4/3}~~~	Q=(-2/3)	~~~\bar{u}\nu  \nonumber \\
\end{equation}
\begin{equation}
\hspace{1.3in}~~~~~~~~~~~(3^*, 1)_{-10/3} ~~~~~~	Q=(-5/3)	~~~ \bar{u}e^-   \nonumber \\
\end{equation}
\begin{equation}
\hspace{1.3in} ~~~~~~~~~~~~~~~~~~~~~ (3^*, 3)_{-5/3}~~~	Q=(-5/3,-2/3,+1/3)	~~~\bar{u}e^-,\bar{u}\nu,\bar{d}\nu   \nonumber \\
\end{equation}

\bigskip

\begin{equation}
B = +1/3, L=-1, ~~~~~~~~~~~~~~~~~~~2(3, 1)_{4/3}~~~	Q=(2/3)	~~~ e^+d     \nonumber \\
\end{equation}
\begin{equation}
\hspace{1.3in}~~~(3, 1)_{10/3} ~~~~~~	Q=(5/3)	~~~ \bar{u}e^-     \nonumber \\
\end{equation}
\begin{equation}
\hspace{1.3in} ~~~~~~~~(3, 3)_{4/3} ~~~~ Q=(-1/3,2/3.5/3) ~~~\nu d,e^+d,e^+u.  \nonumber \\
\end{equation}

\begin{equation}
B = -1/3, L=-1, ~~~~~~2(3^*, 2)_{5/3}~~~	Q=(1/3,4/3)	~~~e^+\bar{u},e^+\bar{d}  \nonumber \\
\end{equation}
\begin{equation}
\hspace{1.3in}~~~(3^*, 2)_{-1/3} ~~~~~~	Q=(-2/3, 1/3)	~~~ \nu\bar{u},e^+\bar{u}   \nonumber \\
\end{equation}

\noindent
The adjoint describes the spin-one gauge bosons of $SU(15)$ and also a spin-zero Higgs
if it is used \cite{FK} for symmetry  breaking. 
\noindent
A spin-one hypothesis would imply that a leptoquark is a gauge boson of $SU(15)$.
In that case, if at least the first two families are treated sequentially as 15's, unless there
is an {\it ad hoc} assumption motivated by the data\cite{CFFIN}, muon-electron LFU 
= Lepton Flavour Universality, meaning that the leptons $e, \mu$ have
identical properties in every way except for their different masses, will be
an inevitable consequence.
\noindent
A spin-zero hypothesis would imply bifermions in the product $15 \times 15 = 105_A + 120_S$ as per their Yukawa interactions, hence we examine the decompositions of 15, 105 and 120 into their SM components, which is easily done with the  Mathematica package LieART  \cite{Feger:2012bs,Feger:2019tvk}:

\begin{equation}
15 = (3, 2)_{+\frac{1}{3}} + (3^*, 1)_{-\frac{4}{3}} + (3^*, 1)_{+\frac{2}{3}} + (1, 2)_{-1} + (1, 1)_{+2}
\end{equation}

\begin{eqnarray}
105 & = & (3^*,2)_{-\frac{1}{3}}+(3, 1)_{+\frac{4}{3}} + (1, 1)_{-2} \nonumber  \\
&&+(3^*,3)_{+\frac{2}{3}}+ (1, 2)_{-1} + (3^*, 1)_{+\frac{2}{3}} + (3, 1)_{-\frac{8}{3}}    \nonumber \\
&&+ (3, 2)_{+\frac{7}{3}} +(6,1)_{+\frac{2}{3}}+ (8, 2)_{-1}             \nonumber \\
&&+(6^*,1)_{-\frac{2}{3}} +(3^*,2)_{-\frac{7}{3}}+ (3^*, 1)_{+\frac{8}{3}} + (3, 1)_{-\frac{2}{3}}+(3,3)_{-\frac{2}{3}}+ (1, 2)_{+1} \nonumber \\
&&+ (8, 2)_{_1} +(3, 1)_{-\frac{2}{3}}+ (1, 2)_{+1}\nonumber \\
\end{eqnarray}

\noindent
and

\begin{eqnarray}
120 & = &  ~~ (6^*,1)_{+\frac{4}{3}}+ (3^*,2)_{-\frac{1}{3}}+(1,3)_{-2}\nonumber \\
&&+(3^*, 1)_{+\frac{2}{3}} +(1,2)_{-1} \nonumber \\
&&+(1,1)_6+(3^*, 1)_{+\frac{2}{3}}+(3,2)_{+\frac{7}{3}}+(6^*, 1)_{-\frac{8}{3}}+(6,3)_{+\frac{2}{3}}+(8,2_{-1})\nonumber \\
&&+(6^*,1)_{-\frac{2}{3}} +(3^*,2)_{-\frac{7}{3}}+ (3^*, 1)_{+\frac{8}{3}} + (3, 1)_{-\frac{2}{3}}+(3,3)_{-\frac{2}{3}}+ (1, 2)_{+1} \nonumber \\
&&+ (8, 2)_{_1} +(3, 1)_{-\frac{2}{3}}+ (1, 2)_{+1}\nonumber \\
\end{eqnarray}
\noindent
The leptoquark $(3^*, 1)_{+\frac{2}{3}}$ which could have fit the now non-existent B anomalies is
seen in both 105 and 120.
Being a weak singlet, it doesn't contribute to the oblique parameters~\cite{Peskin:1991sw} that are tightly constrained by electroweak precision data.
\noindent
The one disadvantage of $SU(15)$, but only an aesthetic one and a stumbling
block we must initially ignore, is that
anomaly cancellation requires the addition of mirror fermions. An advantage
of $SU(15)$ is the absence of proton decay because all of the adjoint
components have well-defined $B$ and $L$ quantum numbers.
\noindent
Even if one rejects the $SU(15)$ model for being vector-like, it is still an ideal testing ground
and classification system of leptoquarks, diquarks and dileptons. {\it i.e.}, it is a perfect umbrella model 
for models with incomplete lists of bifermions.
\noindent
Smoking guns for $SU(15)$ include a predicted enhancement for 
$B \rightarrow K^{(*)}\nu \bar{\nu}$. Because of the lepton
mass dependence in the Higgs Yukawas, it predicts significant
LFU-violating enhancements
relative to the SM for the decays $B^+\rightarrow K^+ \tau^+\tau^-$
and $B_s \rightarrow \tau^+ \tau^-$.
\noindent
In an ingenious argument\cite{GGK}, it has been convincingly shown
that violation of LFU implies the occurrence of LFV decays which are
vanishing in the standard model. These will include the decays
$\tau\rightarrow\mu\gamma$, $\tau \rightarrow\mu\phi$ and 
$B_s \rightarrow \tau\mu$. The discovery of such LFV processes could 
lend support for the additional particles we have discussed.
\noindent
It will be exciting to learn from experiments about more
violations of LFU, as well as any examples of LFV. Such additional
input is necessary to further  evolve the theory.
\noindent
There has been extensive discussion of leptoquarks because they were
temporarily suggested by the now-defunct B anomalies. Bileptons
are suggested by the 331-model. We are tempted to believe that the
third and last type of bifermion, the biquark, appearing in the 224
of $SU(15)$ may also exist in Nature.

\bigskip

\noindent
The 224 has 76 components with $B = L = 0$. The  remaining 148 include
the 72 leptoquarks listed {\it ut supra}, 72 biquarks and 4 bileptons.

\bigskip

\noindent
The 72 biquarks fall into two sets of 36:
\begin{equation}
B = +2/3, L=0, ~~~~~~~~~~~~~~~~~~~(3^*+6, 2)_{5/3}~~~	Q=(1/3,4/3)	~~~uu,dd     \nonumber \\
\end{equation}
\begin{equation}
\hspace{1.3in}~~~(3^*+6, 2)_{1/3} ~~~~~~	Q=(1/3. -2/3)	~~~ ud, dd     \nonumber \\
\end{equation}
and

\begin{equation}
B = -2/3, L=+0, ~~~~~~~~~~~~~(3+6^*, 2)_{-5/3}~~~	Q=(-4/3,-1/3)	~~~\bar{u}\bar{u},\bar{u}\bar{d}   \nonumber \\
\end{equation}
\begin{equation}
\hspace{1.3in}~~~(3+ 6^*, 2)_{1/3} ~~~~~~	Q=(-1/3,  2/3)	~~~ \bar{u}\bar{d}, \bar{d}\bar{d}     \nonumber \\
\end{equation}

\bigskip

\noindent
In the phenomenological analysis of tetraquarks (first discovered in 2003)
and pentaquarks (2015), the name ``diquark" is used for two quarks behaving
together like a molecule, so a diquark is definitely a bound state and not an
elementary particle like a biquark. At present the study of tetraquarks and
pentaquarks is successful\cite{Maiani} by using only diquarks without biquarks.
\bigskip

\noindent
It will be interesting to discover whether biquarks become necessary in these
analyses. The distinction between diquark and biquark could be made
using the same criterion as used in \cite{Weinberg} to decide whether
the deuteron is a bound state or elementary. 

\bigskip

\noindent
Finally, we discuss the four bileptons in the 224 which are in two
$SU(2)$ doublets $(Y^{--},Y^-)$ with $B=0, L=2$, and $(Y^{++},Y^+)$
with $B=0, L=-2$. In the context of the 331-model, they lead \cite{331} to the prediction
of a resonance in same-sign leptons with mass between 1 TeV and 4 TeV, and
width $\Gamma_Y \simeq 0.05-0.10$ TeV.
\noindent
The bilepton resonance in $\mu^{\pm}\mu^{\pm}$ has been the subject of searches
by the\\ ATLAS and CMS Collaborations at the LHC. In 
March 2022, ATLAS published an inconclusive result\cite{ATLAS} about the existence
of the bilepton, putting only a lower mass limit $M_Y > 1.08$ TeV. CMS may have
better momentum resolution and charge identification than ATLAS and may
 therefore be able to investigate the bilepton resonance proper. At the time of writing, CMS
began an in earnest search in October 2022 which is expected to be unblinded 
at some time in 2023.
\noindent
Of the three classes of elementary bifermion (biquark, leptoquark, bilepton)
the one which appears nearest to confirmation at the present time is the bilepton.

\centerline{\bf Acknowledgements} 
The work of C. C. and R. T. is funded by the European Union, Next Generation EU, PNRR project "National Centre for HPC, Big Data and Quantum Computing", project code CN00000013 and by INFN iniziativa specifica QFT-HEP.

\end{document}